# The topological-crystalline-insulator (Pb,Sn)Te - surface states and their spin-polarization

S. Safaei,[1] P. Kacman,[1] and R. Buczko[1, *]

[1]*Institute of Physics, Polish Academy of Sciences, Aleja Lotników 32/46, 02-668 Warsaw, Poland*
(Dated: March 27, 2013)

Using a tight-binding approach we study theoretically the nature of surface states in $Pb_{0.4}Sn_{0.6}Te$ - the newly discovered topological-crystalline-insulator. Apart from the studied before (001) surface states, two other surface families, {011} and {111}, in which the mirror symmetry of the crystal's rock-salt structure plays the same role in topological protection, are considered. Our calculations show that while in (111) surface states of (Pb,Sn)Te four single topologically protected Dirac-cones should appear, for the (110) surface states the protection is lifted for two L points. In this case, instead of the Dirac points energy gaps occur in the surface states, due to the interaction between the two L valleys. In all studied cases a chiral spin texture is obtained.

## I. INTRODUCTION

In the recent years the study of topological phenomena became one of the major topics of condensed-matter physics,.[1,2] This was initiated by the theoretical prediction[3–5] and the successive experimental discovery[6–8] of topological insulators (TIs). In TIs the bulk insulating states are accompanied by metallic helical Dirac-like electronic states on the surface of the crystal. In this class of materials spin-orbit coupling and time-reversal symmetry combine to form topologically protected states and feature a chiral spin texture, thus providing robust spin-polarized conduction channels. In the search for TIs primarily narrow-gap semiconductors, in which the band gaps are smaller than relativistic corrections to the band structure, were considered. The TI phase was first demonstrated in two-dimensions, i.e., in HgTe/HgCdTe quantum wells[6] and also in InAs/GaSb heterostructures.[9] The examples of three-dimensional TI phase include $Bi_2Se_3$ and $Bi_2Te_3$[1,2] with an odd number of band inversions, which support gapless Dirac-like surface states. The narrow gap IV-VI semiconductors PbTe, PbSe and SnTe, as well as their substitutional solid solutions, $Pb_{1-x}Sn_xTe$ and $Pb_{1-x}Sn_xSe$, have been also considered, but they were identified as trivial insulators,[4] because in these the band inversion happens simultaneously at even number (four) of L points in the Brillouin zone. It was suggested that applying uniaxial strain[4] or exploiting the anisotropic energy quantization of electrons confined at an PbTe/(Pb,Sn)Te interface[10] can be used to solve this problem. Next, Liang Fu introduced the notion of "topological crystalline insulators" (TCI)[11] and predicted that in a given class of materials the gapless (001) surface states are supported by fourfold (C4) or sixfold (C6) rotational symmetry. It has been proposed lately, that this novel TCI phase should exist in SnTe.[12] Soon after it has been confirmed by angle-resolved photoelectron spectroscopy (ARPES) studies that indeed metallic surface states exist on the (001) surfaces of SnTe.[13] They have been observed also on the (001) surfaces of the IV-VI substitutional solid solutions $Pb_{1-x}Sn_xSe$[14] and $Pb_{1-x}Sn_xTe$.[15] It has been shown[16] that the compositional disorder in these alloys does not destroy the TCI phase. Very recently, the observation of chiral spin textures of the metallic (001) surface states in the TCI phase of $Pb_{0.73}Sn_{0.27}Se$[17] as well as of $Pb_{0.6}Sn_{0.4}Te$,[15] by spin-resolved photoelectron spectroscopy (SRPES), has been reported. It should be emphasized that the metallic surface states observed in real SnTe- or SnSe-based compounds have almost linear, Dirac-like dispersions,[12–15] due to the spin-orbit interactions.

In this article we present a systematic theoretical study of the electronic structure, in particular the nature of surface states, in $Pb_{0.4}Sn_{0.6}Te$. The Sn content $x = 0.6$ assures the band inversion and the TCI phase in the $Pb_{1-x}Sn_xTe$ material. In this rock-salt TCI the surface states with nontrivial Dirac-like energy spectrum can form at various surfaces of the crystal. Each Dirac point, if it appears, corresponds to one of the four L-points in the three-dimensional bulk Brillouin zone (3DBZ). As shown in ref. 12, these Dirac points are topologically protected only at crystal surfaces symmetric about any of {110} mirror planes (the Dirac points must be situated at the line of such plane symmetry). These are {n n m} surfaces. We study thus, apart from the studied before (001)-oriented surface, the surface states for the (110) and (111) planes. In Fig. 1 the 3DBZs are appropriately oriented to show the projections onto (001), (110) and (111) surfaces. The corresponding {110} mirror planes are marked in yellow. As one can see in Fig. 1, while for the (001)-oriented surface there are two such {110} mirror planes, for the (110) surface only one and for the (111)-oriented there are three such planes. We note that in all three cases the projections of all but one high symmetry L-points of the 3DBZ are situated at the edges of the corresponding two-dimensional Brillouin zones (2DBZ), marked in green in the Figure. The one exception is the $L_1$ point in the (111)-oriented 3DBZ, which projects to the $\bar{\Gamma}$ point in the center of the 2DBZ. We note also that while the 2DBZ for (111) surface is a regular hexagon, for the two other cases the 2DBZs are rectangular. It can be observed that a hexagonal 2DBZ occurs always when the {n n m} surface indexes are all odd numbers. The rectangular shape of the 2DBZ requires different parities of n and m.

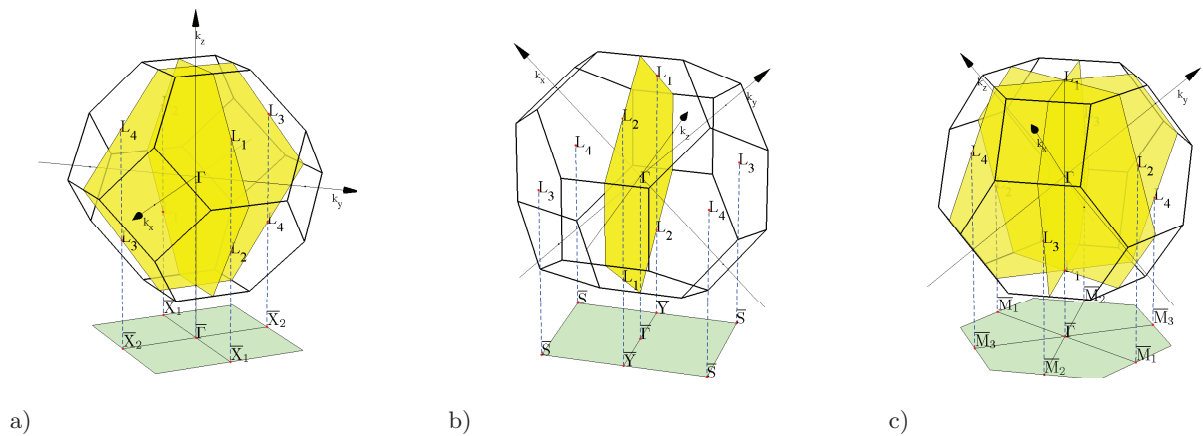

FIG. 1: (Color online) The Brillouin zone for the (a) (001)-, (b) (110)- and (c) (111)-oriented rock-salt crystal with the corresponding 2DBZs (in green). The {110} mirror planes of the (001) (a) (110) (b) or (c) (111) surface are marked in yellow. In the 2DBZs the {110} mirror plane symmetry lines are also depicted.

The spin polarization of metallic surface states in the TCI phase of $Pb_{0.4}Sn_{0.6}Te$ is studied by calculating the in-plane spin texture of the surface states. When the band structure is inverted, chiral spin texture of the states at all considered surfaces, (001), (110) and (111), is anticipated.

## II. BAND-STRUCTURE CALCULATIONS

As mentioned above, successive substitution of Pb by Sn in PbTe strongly changes the relativistic effects and results in a compositional evolution of the band structure of the $Pb_{1-x}Sn_x$Te solid solution. At $x \simeq 0.37$ a band inversion between the topmost valence band and the lowest conduction band occurs, leading to a topological phase transition from a trivial insulator to a TCI state. For higher Sn contents, in the inverted band gap state, the formation of Dirac-like surface states that cross the band gap are expected. To study these effects for different surfaces, we consider an alloy well above the critical composition, i.e., $Pb_{0.4}Sn_{0.6}$Te. The electronic surface states in the substitutional alloy $Pb_{0.4}Sn_{0.6}$Te were obtained by using tight-binding approach and virtual crystal approximation. The parameters for the constituent compounds, PbTe and SnTe, are needed to describe $Pb_{0.4}Sn_{0.6}$Te within the virtual crystal approximation. The tight-binding parameters for both PbTe and SnTe were taken from ref. 18, where they were obtained within a nearest-neighbor 18-orbital $sp^3d^5$ model. Therefore, in our tight-binding Hamiltonian for $Pb_{0.4}Sn_{0.6}$Te, the s, p and d orbitals and nearest-neighbor interactions are included. In all calculations periodic boundary conditions have been imposed in the two directions parallel to the surface.

The (111)-oriented $Pb_{0.4}Sn_{0.6}$Te alloy crystal slab, which has been used in the calculations, consist of 451 atomic monolayers, while the (110)-oriented slab is 315 monolayer-thick. The presented previously in ref. 14 results for the (001)-surface states in $Pb_{0.4}Sn_{0.6}$Te have been obtained by the same method on a slab with 280 monolayers. It should be noted that despite different number of monolayers the thickness of the slabs is similar for all three orientations and is in between 70-90 Å. This results from different inter-layer distances along these directions. The odd number of atomic monolayers in (111) slab enables to study separately the surfaces consisting either of cations or anions. All considered $Pb_{0.4}Sn_{0.6}$Te slabs have rock-salt crystal structure, typical for their component IV-VI compounds.

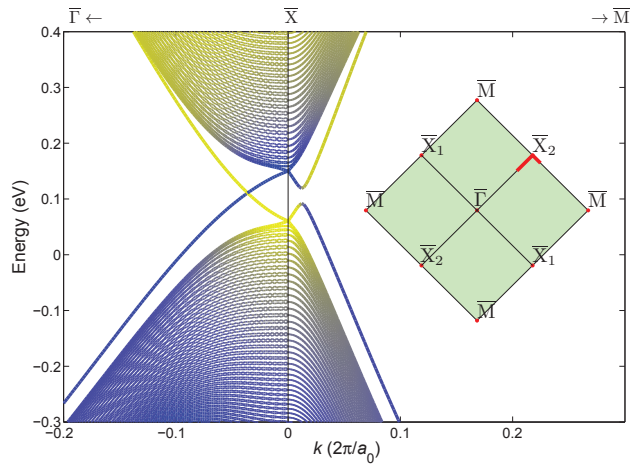

FIG. 2: (Color online) The calculated band structure of a 280-monolayer-thick (001)-oriented $Pb_{0.4}Sn_{0.6}$Te slab. The lines color changes from yellow to blue depending on the cation (yellow) and anion (blue) p-type orbitals dominant contribution to the wavefunction. The green rectangular in the inset is the (001) 2DBZ. The energies are calculated for the k-values shown in red in the inset.

Fig. 2 shows the calculated band structure of (001)-oriented $Pb_{0.4}Sn_{0.6}$Te slab. The k=0 value corresponds to the projection of the $L_3$ and $L_4$ points in the bulk crystal (compare Fig. 1 (a)) onto the (001) surface, i.e., to the $\overline{X}$ point of the surface Brillouin zone. In the Figure the wave vector of electrons $k$ is given in the units $2\pi/a_0$, where $a_0$ is the rock-salt lattice parameter. The energies are calculated for the k-values in the vicinity of the $\overline{X}$ point of the 2DBZ, along the $\overline{\Gamma} - \overline{X} - \overline{M}$ pass, shown by the red line in the inset. The lowest and highest states of the conduction and valence bands, respectively, represent states localized at the surface. For this surface our calculations confirm the findings reported previously – analogous to the situation in SnTe[12] and $Pb_{1-x}Sn_xSe$,[14,17] the (001) surface states for $Pb_{0.4}Sn_{0.6}Te$ are also found to cross the ca 100 meV inverted bulk band gap along the $\overline{\Gamma}$-$\overline{X}$ direction, i.e., they form a Dirac cone. Around all four $\overline{X}$ points presented in the inset, which are equivalent in pairs, the band structure is of course the same as presented in Fig. 2. Due to the interaction between the L valleys, the four Dirac points are not situated at the two $\overline{X}$ points, but are moved along the projections of the {110} mirror planes, i.e., along the $\overline{\Gamma} - \overline{X}$ lines.

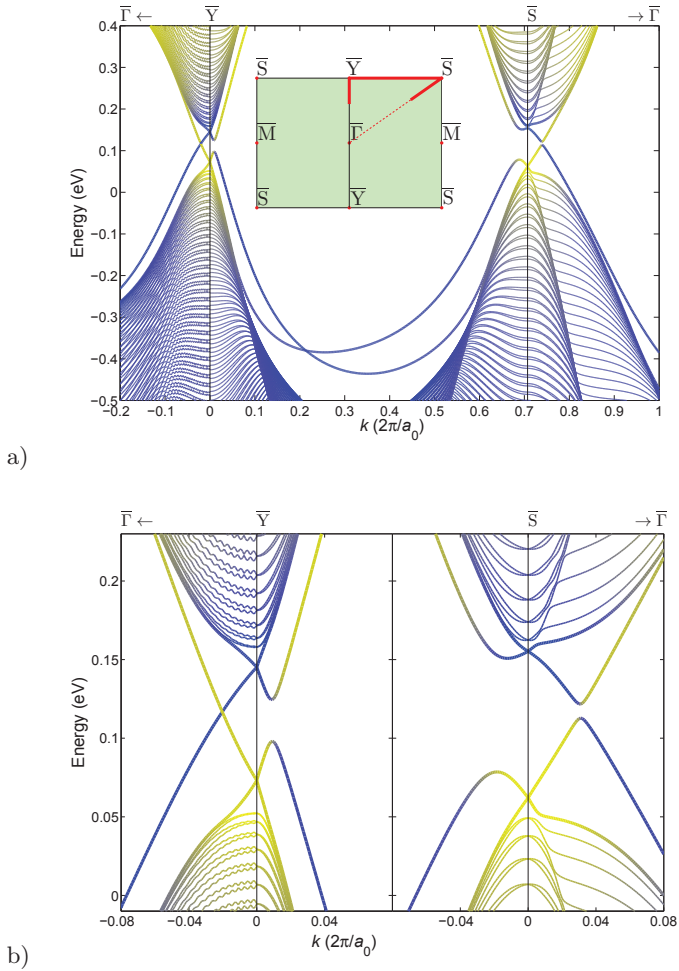

b)

FIG. 3: (Color online) (a) The calculated band structure of a (110)-oriented $Pb_{0.4}Sn_{0.6}Te$ slab, 315 monolayer thick, for the $k$ wave vectors shown in the inset by the red line. The zoomed view (b) of the band structure in the vicinity of the $\overline{Y}$ and $\overline{S}$ points of the 2DBZ has been obtained for a thicker $Pb_{0.4}Sn_{0.6}Te$ slab (365-monolayers).

In Fig. 3 the calculated band structure of a (110)-oriented $Pb_{0.4}Sn_{0.6}Te$ crystal for the $k$ wave vectors along the $\overline{\Gamma} - \overline{Y} - \overline{S} - \overline{\Gamma}$ pass is presented. In part (b) of the Figure the results of a more precise calculation performed on a thicker slab in the vicinity of $\overline{Y}$ and $\overline{S}$ points are shown. One can see that around the $\overline{Y}$ point the bands structure is similar to that for the (001) surface around the $\overline{X}$ point and the surface states crossing in the band gap is observed. We note that the L-points projected to $\overline{Y}$ are the $L_1$ and $L_2$ high symmetry points of the 3DBZ, as presented in Fig. 1 (b). As one can see, these two L-points are situated on the $(\overline{1}10)$ mirror plane of the (110)-surface (marked in yellow in the figure) and thus one can expect the corresponding Dirac points to be topologically protected. At $\overline{S}$ point of the 2DBZ, however, the protection of the (110) surface states is lifted, because the $L_3$ and $L_4$, which are projected to $\overline{S}$ point are not situated in the $(\overline{1}10)$ mirror plane, as shown in Fig. 1 (b). In the latter case, due to the interaction between the two L valleys, instead of Dirac points energy gaps occur in the surface states along the $\overline{S} - \overline{\Gamma}$ lines.

For the last considered surface family, {111}, two cases have to be distinguished, because in this direction the crystal has surfaces composed exclusively of either cations or anions. For both these situations four single, topologically protected Dirac-cones (one in $\overline{\Gamma}$ point and three in the $\overline{M}$ points) are obtained in the calculations. For the anion-terminated slab the bands are brought to contact forming anion Dirac cones, while in the other case the bands meet to form cation Dirac cones. In Fig. 4 the band structure along the $\overline{\Gamma} - \overline{M}$ line, with one pair of the Dirac cones is presented for surfaces consisting of cations (a) and anions (b). For (111) cation surface states the Dirac points appear in the energy gap close to the top of the valence band (see Fig.4 (a)). For the anion surface states (Fig. 4 (b)) the Dirac points are situated just below the conduction band. It should be emphasized that here, in contrast to the described above situation for (001) and (110) surfaces, all the Dirac points are well separated and appear exactly at the appropriate projection points of the single $L$ points. Moreover, as shown in Fig.1 (c), for the (111) surface all $L$ points belong to the three {110} mirror planes of the (111)-surface and, thus, all Dirac points should be topologically protected. We should note, however, that in our study the polarity of the cation/anion-terminated surfaces is not taken into account. The extra confinement related to the charges on the surfaces can lead to additional Rashba splittings, as shown, for example, for $Bi_2Se_3$ by Bahramy et al..[19]

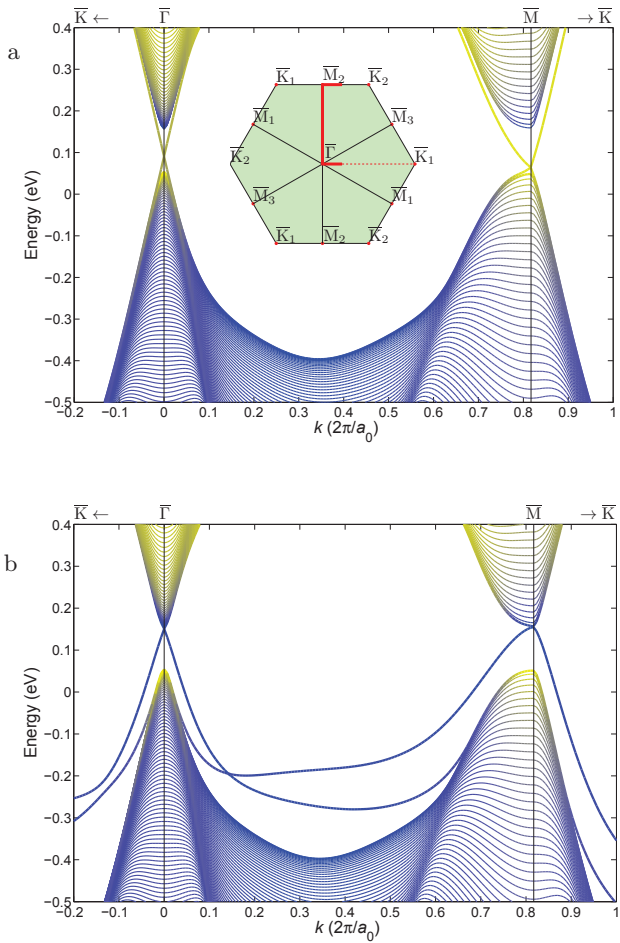

FIG. 4: (Color online) The calculated band structure of a 451-monolayer-thick (111)-oriented $Pb_{0.4}Sn_{0.6}Te$ slab for the $k$ wave vectors of the 2DBZ along the $\overline{K} - \overline{\Gamma} - \overline{M} - \overline{K}$ pass, as shown in the inset by the red line. The band structure of the slab with cations at the surfaces is presented in (a), with anion surfaces in (b). The blue to yellow color coding indicates again the contributions of the cation (yellow) and anion (blue) p-orbitals to the wavefunctions.

## III. SPIN TEXTURE OF THE SURFACE STATES

The calculated spin textures for the states above and below the Dirac points are presented in Fig. 5. These states are spin-polarized and have a chiral spin texture for all three, (001), (110) and (111), orientations of the surfaces. For energies above the Dirac points of the (001)-oriented surface, for wave vectors $k$ between the Dirac points (inner vortex) the spin rotates counter-clockwise about $\overline{X}$. Outside this region (outer vortex) the rotation is reversed (clockwise). In the local environment of each Dirac point small left-handed chiral structure is obtained, like predicted in Ref. 12 and observed in the case of $Pb_{0.6}Sn_{0.4}Te$.[15] For energies below the Dirac point all the chiralities are reversed as compared to those obtained for the band above the Dirac point. The clockwise (counter-clockwise) chirality is related to the cation (anion) p-type orbitals.

For the (110) surface states the spin pattern around the $\overline{Y}$ projection point has also a multi-vortical structure, similar to the one for (001) surface states around the $\overline{X}$ point. Again, the rotation in the outer vortex is opposite to that in the inner vortex. Although a multi-vortical structure is also obtained for (110) surface states around the $\overline{S}$ projection, the spin texture in this case looks much more complicated. Here, instead of two local vortexes around the Dirac points, four local vortexes along the $\overline{S} - \overline{\Gamma}$ lines are obtained. These local vortexes are related to the four energy gaps in the surface states.

For the (111) anion surface states the spin texture is presented only below the Dirac point, because above the Dirac point the surface states are degenerated with the conduction band. Similarly, the cation surface states below the Dirac point enter the valence band. In Fig. 5 we present, therefore, the contour plots for the cation surface states only above the Dirac point. Both these plots have the form of single vortexes. The spins in the anion(cation) Dirac lower(upper) cones are rotating in the opposite directions. We note that these chiralities are the same as those predicted in Ref. 20 and observed in 3D TIs $Bi_2Se_3$ and $Bi_2Te_3$.[8] The same chirality was also predicted in the case of PbTe/(Pb,Sn)Te (111)-interface.[10]

Finally, we have also checked that in the (001) and (110) cases the perpendicular to the surface spin component is equal to zero in the whole region shown in Fig. 5. For (111) surface, while in the close vicinity of the Dirac point in $\overline{\Gamma}$ it also equals zero, in the Dirac cones around the $\overline{M}$ points the out-of-surface spin component appears, because here the L-valleys' directions are not perpendicular to the surface. Again, the same different behavior of the out-of-(111)-surface spin component in the $\overline{\Gamma}$ and $\overline{M}$ points has been also predicted for $Bi_2Se_3$[20] and the PbTe/(Pb,Sn)Te interface.[10]

## IV. DISCUSSION AND CONCLUSIONS

In general, for a {n n m} surface only one mirror plane of {110} type can be found. This leads to only one pair of protected Dirac points, which correspond to the two L-points situated at this mirror plane in the 3DBZ. Two special cases can be distinguished: n=m=1 and n=0. While for the former case three {110} mirror planes exist, for the latter case there are two such planes (compare Fig. 1). We note that despite different number of mirror planes, the number of Dirac points should be the same for {001} and {111} surfaces, because for the latter one L-point is common for all three mirror planes. Therefore, in both these cases four Dirac points should appear in the band structure. In both these cases all L-points are located at the symmetry planes and the corresponding Dirac points should be topologically protected.

As shown in the Appendix, for any {n k m} surface the four L-points in the 3DBZ project to four different points of the 2DBZ only when all $n, k$ and $m$ surface indexes have the same parity. Otherwise they project in pairs. Thus, when the parities of n and m in a {n n m} surface are different, the L-points are projected in pairs. In general, only one pair of Dirac points is guaranteed to appear on the mirror symmetry line in the vicinity of the projection of one L points pair and is topologically protected. However, for the special case of {001} surfaces there are two mirror symmetry lines and all four Dirac points are topologically protected. Two of them are shown in Fig. 6(a). Similar band structure is obtained for the (110) surface in the vicinity of the $\overline{Y}$ point of the 2DBZ. In this case only one pair of Dirac points exists - for the L-points projected to $\overline{S}$ instead of Dirac points gaps occur for the (110) surface states, as shown in Fig. 6(b).

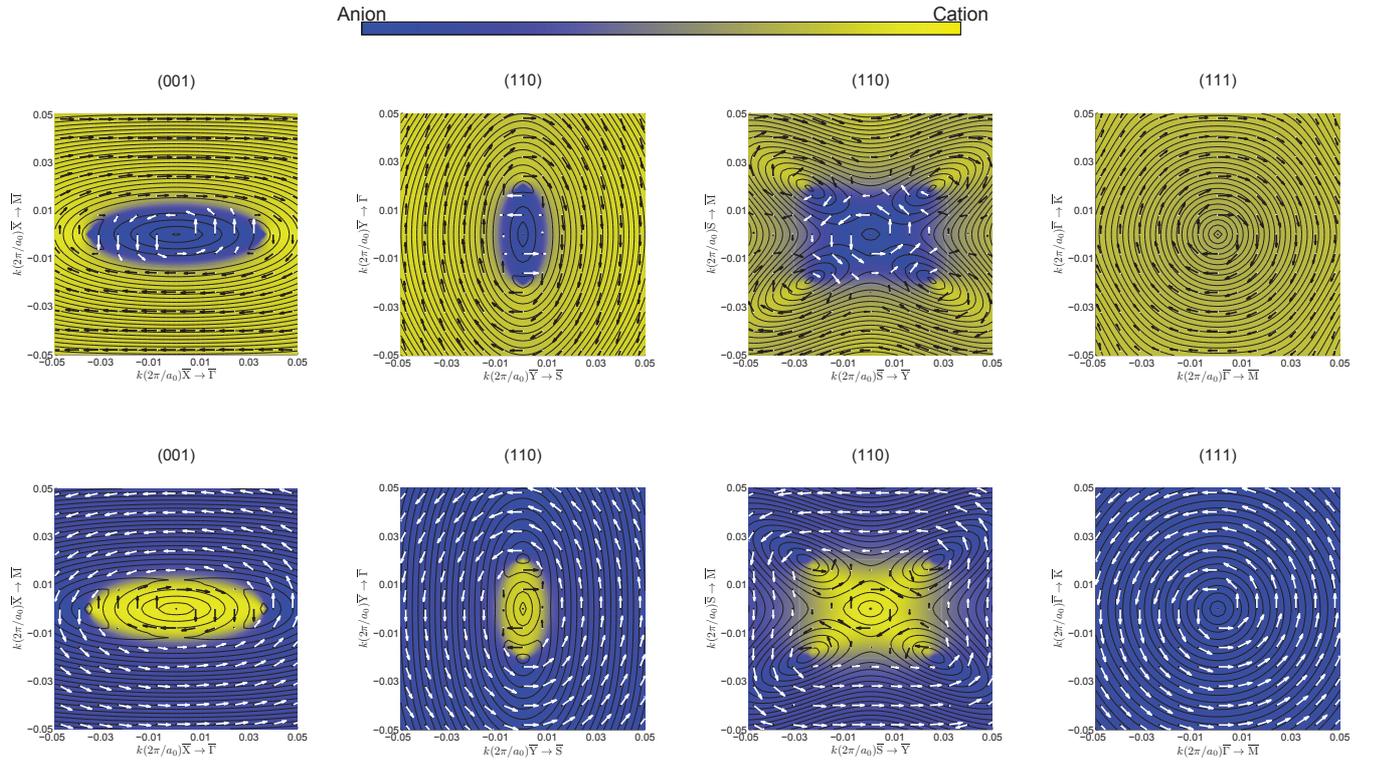

FIG. 5: (Color online) Contour plots of the constant-energy lines of the (001), (110) and (111) surface states of $Pb_{0.4}Sn_{0.6}Te$ above (upper row) and below (lower row) the Dirac point. For the (001) surface the plots are around the $\overline{X}$ point of the 2DBZ, for (110) around two, $\overline{Y}$ (left) and $\overline{S}$ (right), points and for (111) in the vicinity of $\overline{\Gamma}$. For the (111)-oriented surface, in the upper panel the result obtained for the cation-ended slab is shown, while the lower panel shows the result for the anion-ended slab. The arrows indicate the in-plane spin texture, the arrows' size the degree of spin polarization. The blue to yellow color coding indicates again the contributions of the cation (yellow) and anion (blue) p-orbitals to the wavefunctions, as shown in the bar.

In the {n n m} case the L-points project to different points in the 2DBZ only when n and m have the same parity (it means of course that they are both odd numbers).

In Fig. 7 the band structures around $\overline{\Gamma}$ and around $\overline{M}$ points obtained for the (111)-oriented cation surface are presented (analogous results are obtained for the (111)-oriented slab with anions at the surfaces). While at the $\overline{\Gamma}$ point isotropic Dirac-cone is observed, the band structure around $\overline{M}$ is strongly anisotropic, i.e., along $\overline{M} - \overline{\Gamma}$ line depends differently on the $k$ values than in $\overline{M} - \overline{K}$ direction. The difference in shapes of the band structures is due to different orientations of the constant energy ellipsoids around different L points. While the $L_1$ ellipsoid is projected along its long axis, the ellipsoids for the other three L points are tilted to the projection direction.

Finally, chiral spin texture of the surface states of $Pb_{0.4}Sn_{0.6}Te$ for all considered surfaces, (001), (110) and (111), has been obtained. While our calculations show a "multi-vortical" spin structure for the (001) and (110) surface states, for the (111) anion and cation surfaces a "single-vortical" spin texture is anticipated. As revealed by our studies, the spin polarization seems to be inherent to all (001), (110) and (111) surface states in narrow-gap IV-VI semiconductors in their TCI phase. We note that in case of (001) and (110) surfaces the topology of the constant energy lines undergoes a Lifshitz transition (compare Fig. 5), similar to that predicted and observed for SnTe.[12,13] It seems that such transition has been also observed in $Pb_{1-x}Sn_xTe$.[15]

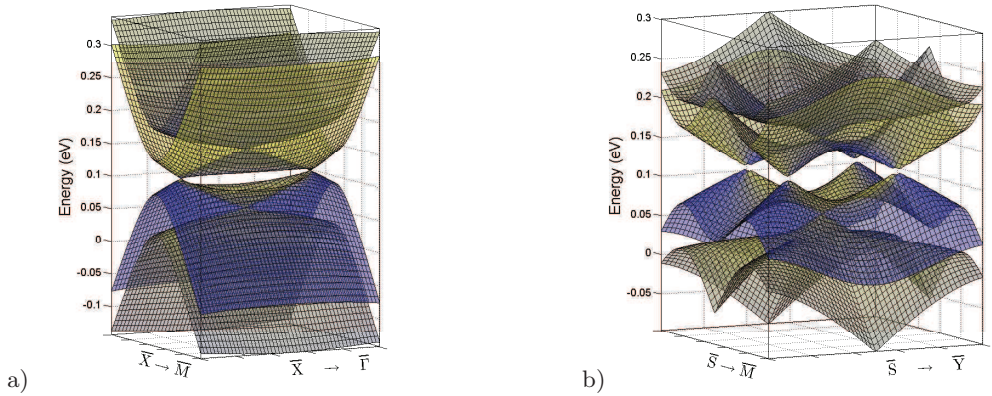

FIG. 6: (Color online) The three-dimensional view of the band structure in the vicinity of the $\overline{X}$ for the (001)-oriented surface (a) and in the vicinity of $\overline{S}$ point of the 2DBZ for the (110)-oriented surface (b) of $Pb_{0.4}Sn_{0.6}Te$.

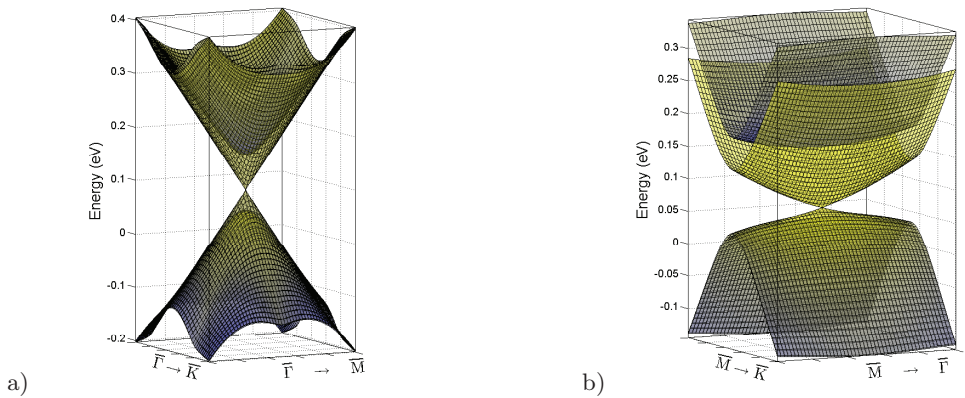

FIG. 7: (Color online) The three dimensional view of the band structure in the vicinity of the $\overline{\Gamma}$ (a) and $\overline{M}$ (b) points of the 2DBZ for the (111)-oriented cation surface of $Pb_{0.4}Sn_{0.6}Te$.

Qualitatively similar effects to those described above we expect also to occur for the states of the (001)-, (110)- and (111)-oriented interfaces between TCI and a normal insulator. For the (111) interface, for example, this has been already shown in Ref. 10. However, all the experiments for the TCI phase in PbTe-SnTe solid solutions have been performed on (001) surfaces, because, as it is well known, the (001) surfaces are the cleavage planes in the $Pb_{1-x}Sn_xTe$ material. Still, it is also well known that using the $BaF_2$ substrate allows the growth of Se- and Te-based rock-salt IV-VI compound crystals and heterostructures in the [111] direction and that (111)-oriented IV-VI structures have the highest mobilities and electron mean free paths.[21] These make the (111) surface states most interesting for the study of TCI and other transport phenomena. It was also shown that using the GaAs substrate with a thick CdTe buffer allows growth by molecular beam epitaxy of good quality PbTe-based structures.[22] As the (110)-oriented GaAs substrates are available, it seems that by the same method the growth of the IV-VI structures along the [110] crystallographic axis should also be possible. We are convinced that our study of the surface states for $Pb_{1-x}Sn_xTe$ TCI crystal oriented in these directions can be very useful for predicting/describing various phenomena in such structures.

## V. APPENDIX - RULES FOR THE L-POINTS PROJECTION ONTO AN ARBITRARY SURFACE

The study of the L-points projection onto an arbitrary surface is motivated by the presented above results that in the case of (001) and (110) surfaces two pairs of nonequivalent $L$ points are projected onto two nonequivalent points in 2DBZ, whereas for the (111) sur-face four $L$ points are projected onto four nonequivalent points in the appropriate 2DBZ. Therefore, we would like to find a general rule to describe for which surfaces $L$-points are projected separately and for which they are projected in pairs. Let us first define the vectors $\vec{L}_i$ for the four nonequivalent $L$ points in the first Brillouin zone as $\vec{L}_i = \vec{G}_i/2$, $i = 1,..,4$, where $\vec{G}_i$, in units of $2\pi/a_0$, are given by:

$$\vec{G}_1 = \begin{pmatrix} 1 \\ 1 \\ 1 \end{pmatrix}, \quad \vec{G}_2 = \begin{pmatrix} 1 \\ 1 \\ -1 \end{pmatrix},$$
$$\vec{G}_3 = \begin{pmatrix} 1 \\ -1 \\ -1 \end{pmatrix}, \quad \vec{G}_4 = \begin{pmatrix} 1 \\ -1 \\ 1 \end{pmatrix} \quad (1)$$

The first three $\vec{G}_i$ vectors we choose as a basis in the reciprocal space.

One can say that two points $L_i$ and $L_j$ project onto the same point in the 2DBZ, when $L_i$ and $L_j$ (or any other point equivalent to $L_j$) are located on the same line along the projection direction. For a given $s_1\ s_2\ s_3$ surface the projection direction is given by the vector $\vec{S} = \begin{pmatrix} s_1 & s_2 & s_3 \end{pmatrix}^T$. Thus, the two L points project onto the same point of the 2DBZ when:

$$\vec{L}_i - \vec{L}_j = \vec{G} + \alpha \vec{S}, \quad (2)$$

where $\vec{G}$ is the reciprocal net vector $n_1\vec{G}_1 + n_2\vec{G}_2 + n_3\vec{G}_3$ and $n_1, n_2, n_3$ are integer numbers. The condition for the $s_1, s_2, s_3$ surface indexes leading to the projection of $L_i$ and $L_j$ points onto the common point in the 2DBZ will be:

$$\vec{L}_i - \vec{L}_j = \begin{pmatrix} l_1 \\ l_2 \\ l_3 \end{pmatrix} = \begin{pmatrix} n_1 + n_2 + n_3 \\ n_1 + n_2 - n_3 \\ n_1 - n_2 - n_3 \end{pmatrix} + \alpha \begin{pmatrix} s_1 \\ s_2 \\ s_3 \end{pmatrix} \quad (3)$$

We note that $s_1, s_2, s_3$ being the surface indexes are relative primes. Moreover, $l_1, l_2, l_3$ are either $\pm 1$ or 0 and the $l_k$ components for any $L_i$, $L_j$ pair have different parities – at least one $l_k$ is equal 0 and at least one has to be $\pm 1$. In contrast, for any set of $n_k$ ($k = 1,..,3$) all three components of $\vec{G}$ have the same parity. We observe that Eq. 3 can be satisfied only when two conditions are simultaneously fulfilled: (a) $\alpha$ is an odd integer (one can always find a solution with $\alpha = 1$) and (b) the parities of $s_k$ are either the same as the parities of corresponding $l_k$ (with even $\vec{G}$ components) or $s_k$ and $l_k$ have opposite parities (with odd $\vec{G}$ components). Thus, we can conclude that two $L$ points project to the same point on 2DBZ of a $\{s_1, s_2, s_3\}$ surface only when the parities of the $s_k$ indexes are various.

Let us take, for example, two pairs of $L$-points, $(L_1, L_2)$ and $(L_4, L_3)$. For both $\vec{L}_1 - \vec{L}_2$ and $\vec{L}_4 - \vec{L}_3$, $l_1 = l_2 = 0$ and $l_3 = 1$. This means that the condition (2) can be satisfied with the same surface indexes for these two pairs – in this particular case the parity of the $s_3$ index has to be opposite to the parity of $s_1$ and $s_2$. We note that these indexes do not satisfy the condition (2) for $(L_1, L_3)$ pair. Thus, the two, $(L_1, L_2)$ and $(L_3, L_4)$, pairs are projected onto two different, nonequivalent points in 2DBZ.

After examination of all possible $L$-points pairs we can conclude that the following four nonequivalent L-points are projected in two pairs on a $\{s_1, s_2, s_3\}$ surface :
$(L_1, L_2)$ and $(L_3, L_4)$, if $s_3$ parity is opposite to $s_1$ and $s_2$ parities;
$(L_1, L_3)$ and $(L_2, L_4)$, if $s_1$ parity is opposite to $s_2$ and $s_3$ parities;
$(L_1, L_4)$ and $(L_2, L_3)$, if $s_2$ parity is opposite to $s_1$ and $s_3$ parities.

In contrast, when the surface indexes $s_1, s_2, s_3$ have the same parity (they are odd numbers, as indexes are relative primes), the condition (2) can not be satisfied for any pair of $L$ points. Then the four $L$ points are projected onto four separate nonequivalent points in 2DBZ. Moreover, one can prove by similar considerations that in this case one of $L$ points has to be projected onto the $\overline{\Gamma}$ point, in contrast to the other $L$-points, which are projected onto the edges of 2DBZ.

### ACKNOWLEDGMENTS

We would like to thank Tomasz Story and Marta Galicka for their diverse support. Financial support from the EC Network SemiSpinNet (PITN-GA-2008-215368), the European Regional Development Fund through the Innovative Economy grant (POIG.01.01.02-00-108/09), and the Polish National Science Centre (NCN) Grant No. 2011/03/B/ST3/02659 is gratefully acknowledged.

---

* [buczko@ifpan.edu.pl](buczko@ifpan.edu.pl)
[1] M. Z. Hasan and C. L. Kane, Rev. Mod. Phys. **82**, 3045 (2010).
[2] X.-L. Qi and S.-C. Zhang, Rev. Mod. Phys. **83**, 1057 (2011).
[3] C. L. Kane and E. J. Mele, Phys. Rev. Lett. **95**, 146802 (2005).
[4] L. Fu and C. L. Kane, Phys. Rev. B **76**, 045302 (2007).
[5] L. Fu, C. L. Kane, and E. J. Mele, Phys. Rev. Lett. **98**, 106803 (2007).
[6] M. König, S. Wiedmann, C. Brüne, A. Roth, H. Buhmann, L. W. Molenkamp, X.-L. Qi, and S.-C. Zhang, Science **318**, 766 (2007).
[7] D. Hsieh, D. Qian, L. Wray, Y. Xia, Y. S. Hor, R. J. Cava, and M. Z. Hasan, Nature **452**, 970 (2008).

[8] D. Hsieh, Y. Xia, D. Qian, L. Wray, J. H. Dil, F. Meier, J. Osterwalder, L. Patthey, J. G. Checkelsky, N. P. Ong, A. V. Fedorov, H. Lin, A. Bansil, D. Grauer, Y. S. Hor, R. J. Cava, and M. Z. Hasan, Nature 460, 1101 (2009).

[9] I. Knez, R.-R. Du, and G. Sullivan, Phys. Rev. Lett. 107, 136603 (2011).

[10] R. Buczko and L. Cywinski, Phys. Rev. B 85, 205319 (2012).

[11] L. Fu, Phys. Rev. Lett. 106, 106802 (2011).

[12] T. H. Hsieh, H. Lin, J. Liu, W. Duan, A. Bansil, and L. Fu, Nat. Commun. 3, 982 (2012).

[13] Y. Tanaka, Z. Ren, T. Sato, K. Nakayama, S. Souma, T. Takahashi, K. Segawa, and Y. Ando, Nat. Phys. 8, 800 (2012).

[14] P. Dziawa, B. J. Kowalski, K. Dybko, A. Szczerbakow, M. Szot, E. Łusakowska, T. Balasubramanian, B. M. Wojek, M. H. Berntsen, O. Tjernberg, and T. Story, Nat. Mater. 11, 1023 (2012).

[15] S.-Y. Xu, C. Liu, N. Alidoust, M. Neupane, D. Qian, I. Belopolski, J. D. Denlinger, Y. J. Wang, H. Lin, L. A. Wray, G. Landolt, B. Slomski, J. H. Dil, A. Marcinkova, E. Morosan, Q. Gibson, R. Sankar, F. C. Chou, R. J. Cava, A. Bansil, and M. Z. Hasan, Nat. Commun. 3, 1192 (2012).

[16] L. Fu and C. L. Kane, Phys. Rev. Lett. 109, 246605 (2012).

[17] B. M. Wojek, R. Buczko, S. Safaei, P. Dziawa, B. J. Kowalski, M. H. Berntsen, T. Balasubramanian, M. Leandersson, A. Szczerbakow, P. Kacman, T. Story, and O. Tjernberg", Phys. Rev. B 87, 115106 (2013).

[18] C. S. Lent, M. A. Bowen, J. D. Dow, R. S. Allgaier, O. F. Sankey, and E. S. Ho, Superlattices Microstruct. 2, 491 (1986).

[19] M. S. Bahramy, P. King, A. de la Torre, J. Chang, M. Shi, L. Patthey, G. Balakrishnan, P. Hofmann, R. Arita, N. Nagaosa, and F. Baumberger, Nat. Commun. 3, 1159 (2012).

[20] F. Zhang, C. L. Kane, and E. J. Mele, Phys. Rev. B 86, 081303 (2012).

[21] G. Grabecki, K. A. Kolwas, J. Wrobel, K. Kapcia, R. Puzniak, R. Jakiela, M. Aleszkiewicz, T. Dietl, G. Springholz, and G. Bauer, J.Appl.Phys. 108, 053714 (2010).

[22] M. Szot, K. Dybko, P. Dziawa, L. Kowalczyk, E. Smajek, V. Domukhovski, B. Taliashvili, P. Dluzewski, A. Reszka, B. J. Kowalski, M. Wiater, T. Wojtowicz, and T. Story, Cryst. Growth Des. 11, 4794 (2011).